\documentclass[conference]{IEEEtran}
\IEEEoverridecommandlockouts
\usepackage{cite}
\usepackage{amsmath,amssymb,amsfonts}
\usepackage{algorithmic}
\usepackage{graphicx}
\usepackage{textcomp}
\usepackage{xcolor}
\usepackage{url}
\usepackage{float}
\usepackage{subcaption}
\def\BibTeX{{\rm B\kern-.05em{\sc i\kern-.025em b}\kern-.08em
    T\kern-.1667em\lower.7ex\hbox{E}\kern-.125emX}}
\begin{document}

\newcommand{\etal}{\textit{et al.}}
\title{Performance Characterization of dApps in Open Radio Access Networks}





\author{
\IEEEauthorblockN{
Conrado Boeira\IEEEauthorrefmark{1},
Eduardo Baena\IEEEauthorrefmark{2},
Andrea Lacava\IEEEauthorrefmark{2},
Tommaso Melodia\IEEEauthorrefmark{2},
Dimitrios Koutsonikolas\IEEEauthorrefmark{2},
Israat Haque\IEEEauthorrefmark{1}
}

\IEEEauthorblockA{\IEEEauthorrefmark{1}
Dalhousie University, Halifax, Canada \\
Email: \{conrado.boeira, israat\}@dal.ca}
\IEEEauthorblockA{\IEEEauthorrefmark{2}
Northeastern University, Boston, USA \\
Email: \{e.baena, lacava.a, t.melodia, d.koutsonikolas\}@northeastern.edu}
}




\maketitle

\begin{abstract}
Despite recommendations to deploy real-time Open Radio Access Network (O-RAN) applications (dApps) in containerized environments, existing approaches predominantly rely on bare-metal servers. Moreover, current dApp deployments offer limited visibility into the resource usage patterns of both intelligent and non-intelligent dApps, hindering informed deployment decisions. This work addresses these gaps by implementing and evaluating representative dApps across multiple deployment scenarios (bare-metal and containers) to characterize the trade-offs in latency, scalability, and resource utilization. Additionally, we identify key performance bottlenecks and demonstrate how offloading dApps to emerging hardware accelerators, such as smart Network Interface Cards (NICs), can alleviate these limitations and improve real-time responsiveness in O-RAN systems.
\end{abstract}


\newcommand{\findingsbox}[1]{
\begin{tcolorbox}[breakable,width=\linewidth,
boxrule=0pt, leftrule = 6pt, top=1pt, bottom=1pt, left=1pt,right=1pt, 
colback=gray!20,colframe=gray!60]
\textbf{Summary:} #1
\end{tcolorbox}
}

\section{Introduction}
\label{sec:introduction}

Open RAN (O-RAN) disaggregates cellular base stations into programmable components (Radio Units (RUs), Distributed Units (DUs), and Central Units (CUs)) deployed on commodity edge servers.  RUs handle RF transmission/reception, DUs perform baseband processing, and CUs manage control plane functions. 
The RAN Intelligent Controller (RIC) orchestrates these components and hosts network applications with varying latency requirements: rApps ($>$1\,s) and xApps (10-1000\,ms).
For real-time operations ($<$10\,ms), dApps have been proposed~\cite{d2022dapps} as software components integrated within the RAN logical nodes. Unlike Mobile Edge Computing (MEC) applications that consume network services externally, dApps execute \emph{within} the RAN protocol stack, processing raw I/Q samples and returning control decisions to the DU within a single scheduling period. dApps access data unavailable to the RIC such as I/Q samples through the E3 interface to enable use cases such as dynamic spectrum sharing~\cite{gangula2024listen}, intelligent resource allocation, and RF anomaly detection~\cite{scalingi2024det}. As a concrete example, a spectrum-sensing dApp must detect an interferer and instruct the DU to change channels within one scheduling frame ($<$10\,ms).

However, deploying dApps at scale presents a fundamental tension between standards compliance and performance. The O-RAN Alliance mandates containerized microservice  deployments for all RIC applications to ensure portability, fault isolation, and multi-tenancy \cite{o-ran2024dapps}. Specifically, dApps can be co-located with the DU in the same container or deployed in a separate container. The former approach reduces communication latency at the cost of resource (computing and memory) contention, while the separated container-based deployment enables resource and fault isolation at the cost of increased latency. However, \textit{all existing dApp implementations use bare-metal deployments}~\cite{lacava2025dapps,olimpieri2025libiq}, completely bypassing containerization. This creates a critical gap: operators must choose between violating O-RAN specifications or deploying containers without understanding their impact on sub-10\,ms timeline. 

Furthermore, existing works failed to incorporate the emerging programmable smart Network Interface Cards (NICs) as alternative computing platforms of dApps. Smart NICs sit inline between the RU and DU, equipped with their own multi-core processors and memory; this positioning allows potential dApp deployments to intercept I/Q samples directly from fronthaul frames, bypassing the host CPU and the E3 protocol stack. Because dApps share CPU cores and the E3 protocol pipeline with the DU, their overhead sensitivity differs from that of general-purpose containerized microservices, where resource contention is decoupled from the application data path.

Therefore, it is essential to characterize the performance of different dApps in potential deployment scenarios to realize a guideline for informed deployment decision making to guarantee the required latency demand of dApps. The characterization problem can be compounded by three unanswered questions. First, what is the performance cost of O-RAN compliance? Containers introduce virtualization layers (cgroups, network namespaces) that could violate real-time constraints, but no prior work has quantified this overhead for sub-10\,ms workloads. Second, how many concurrent dApps can a single edge server support? As operators move from single-dApp proofs-of-concept to multi-tenant production deployments, understanding scalability limits and resource provisioning requirements becomes essential. Third, can hardware accelerators like smart NICs mitigate performance bottlenecks? While prior work explores GPU acceleration for ML inference \cite{scalingi2024det}, the end-to-end pipeline includes protocol parsing and preprocessing phases that may become the dominant bottleneck.

This paper fills the identified gaps by providing the first systematic evaluation of O-RAN-compliant containerized dApp deployments. We implement four representative workloads spanning the computational complexity spectrum (energy-based spectrum sensing (EBS), Fast Fourier Transform (FFT), fully-connected neural network (FCN), and Xception CNN) in OpenAirInterface \cite{kaltenberger2020openairinterface}. 
We selected these dApps as a way to represent a wide range of computational patterns commonly expected in real-time applications for O-RAN. These include low-complexity statistical processing (EBS), signal-domain transforms (FFT), dense neural networks (FCN), and deep convolutional architectures (Xception CNN). Together, these workloads span increasing levels of computational intensity and model complexity, enabling a comprehensive evaluation of system performance across heterogeneous processing demands.
We evaluate these across four deployment scenarios: (1) bare-metal (non-compliant baseline), (2) co-located containers (dApp+RAN in one container), (3) separated containers (isolated via virtual network interface), and (4) smart NIC offloading. Our testbed comprises a 10-core Intel Xeon server with NVIDIA RTX 8000 GPU and BlueField-3 smart NIC. 

Our contributions are summarized as follows:

\begin{itemize}
    \item First performance characterization of O-RAN-compliant containerized dApps, quantifying the cost of compliance with standards  for real-time applications. This analysis can provide valuable insights for operators seeking to deploy dApps.

    \item Identification of the CPU as the fundamental scalability bottleneck even for GPU-accelerated workloads, with implications for capacity planning. This finding is both non-intuitive and impactful, as it challenges prevailing assumptions regarding GPU-centric performance scaling.

    \item Demonstration of smart NIC datapath-integrated acceleration as a novel offloading paradigm that takes advantage of architectural positioning. Thus revealing an underexplored design space for performance optimization.

    \item Implementation of open-source  and reproducibility artifacts available upon acceptance of the paper~\cite{dapppinetrepo}.

\end{itemize}     

The rest of the paper is organized as follows. Section~\ref{sec:background} provides the necessary background for the work. Section~\ref{sec:related_work} highlights similar works in the field and outlines the gap this work aims to fill. Section~\ref{sec:methodology} describes our setup and the evaluation methodology utilized. Section~\ref{sec:evaluation} presents our results and key take-ways. Finally, Section~\ref{sec:conclusion} concludes the paper.


\section{The Road to dApp Deployment}
\label{sec:background}

This section presents the necessary background and motivation for the proposed work, defining where, why and how dApps are deployed in O-RAN deployments.

\textbf{O-RAN Architecture.}
A typical RAN consists of a collection of base stations equipped with antennas, transceivers, and computing components that enable communication between user equipment (UE) and the core network or the Internet. These base stations are typically closed and vendor-locked, making them difficult to configure or update to meet the rapidly evolving demands of mobile applications. To address these limitations, O-RAN emerged with three key principles: disaggregation of network functions, virtualization of components, and open interfaces for multi-vendor interoperability.  Disaggregation separates the monolithic base station into specialized components: Radio Units (RUs) handle radio frequency operations near antennas, Distributed Units (DUs) manage lower-layer protocols, Centralized Units (CUs) control higher-layer functions, and RAN Intelligent Controllers (RICs) provide programmable network control. This separation enables flexible deployment strategies—DUs and CUs can run on commodity edge servers, while RUs are deployed on specialized hardware (FPGAs or ASICs) near antennas \cite{polese2023understanding}. Furthermore, 3GPP defines multiple functional splits between these components to enable fine-grained resource management. Split 7.2x has become the \textit{de facto} standard, offering optimal balance between performance and complexity, where RUs handle time-sensitive operations (FFT, RF processing), DUs manage physical layer, MAC, and Radio Link Control (RLC) functions, while CUs control higher-layer protocols including Radio Resource Control (RRC) and some RLC functions (optionally). 

In alignment with traditional software-defined networking (SDN) principles, O-RAN components and functionalities are organized into the data plane (RU, DU, CU) and the control plane (e.g., RIC). These two planes communicate through standardized northbound (O1, A1) and southbound (E2) interfaces, enabling a programmable environment for efficient configuration and management. The RIC uses these interfaces to collect Key Performance Indicators (KPIs) from data plane components for data driven decision-making.

\textbf{What are dApps? Why are they crucial for 5G use cases?}
The functionalities of the RIC can be categorized based on their timing requirements into rApps, xApps, and dApps. The rApps operate in the non real-time domain (more than one second) and handle complex, higher-level functions such as network slicing and policy management. In contrast, xApps function within the near real-time domain (approximately 10 to 1000\,ms) to support time-sensitive operations like mobility and handover management. Finally, dApps represent real-time functions that must guarantee an end-to-end latency of no more than 10\,ms. dApps interact with other RAN components through a dedicated, extended interface known as E3, designed to complement the existing E2 interface for xApps. The list of functions envisioned as dApps includes beamforming management, spectrum sensing, and radio frequency fingerprinting. These applications share similar characteristics in terms of the need for fast control loops and the usage of user-plane data (e.g., raw I/Q samples). In order to execute fine-grained tasks on lower layers of the protocol stack, dApps need to respond in real-time. Thus, it is infeasible to move user-plane data around the network; we need to process them close to base stations. Therefore, dApps are deployed close to the DU/CU, instead of the traditional RIC, to limit the communication delay.

dApps execute a closed-loop cycle comprising four phases. Taking spectrum sensing as an example: (1)~\emph{Collection}: the DU encodes raw I/Q samples into an E3 Indication Message and delivers it to the dApp; (2)~\emph{Processing}: the dApp processes the incoming data, for example running ML inference or calculating energy; (3)~\emph{Create Control}: the dApp formulates a control action (e.g., channel change); (4)~\emph{Deliver Control}: the decision is encoded into an E3 Control Message and sent back to the DU. The sum of these four phases plus network round-trip time must remain below 10\,ms. Each phase competes with the DU processing budget on shared CPU cores.

\textbf{What are smart NICs and how can they help?} As dApps operate under strict time constraints, any hardware capable of accelerating and improving their computation warrants investigation. Among these options, one particularly promising yet underexplored candidate is the smart NIC. A smart NIC is a Network Interface Card enhanced with their own processing units, allowing the NIC to perform custom processing of the incoming traffic. Its proximity to the data path allows it to process packets before they traverse the PCIe bus, thus allowing for offloading of applications from the host processing unit to the NIC. Smart NICs usually fit into one of two categories: \textit{on-path} and \textit{off-path}. \textit{On-path} smart NICs will process every packet flowing from the network into the attached host, requiring quite efficient and powerful process units so to not impact the communication latency. \textit{Off-path} smart NICs have their processing units not directly in the data path, allowing for forwarding rules to be created that will either forward traffic to the host or to the smart NIC so it can process then. In this work we will be using the latter, as we have strict latency requirements for communication.

\section{Related Work}
\label{sec:related_work}

This section outlines the existing works focused on performance characterizations in O-RAN, dApps, and similar applications as the ones we explore in this work.


\textbf{O-RAN Performance Characterization.} Recent work has characterized the performance of disaggregated O-RAN components. Wei et al. \cite{wei20225gperf} evaluated the performance of disaggregated O-RAN functionalities using OpenAirInterface, measuring user throughput and latency under varying CPU loads. They observed how the DU CPU utilization tends to be the main bottleneck, with the CU CPU usage remaining relatively low. Similarly, Hervas et al. \cite{hervas2023impact} studied virtual RAN performance under varying CPU allocations and observed that insufficient computational resources degraded base station performance, but increasing CPU beyond an optimal point yielded no further gains and only wasted resources.
Schiavo et al. \cite{schiavo2024yinyangran} evaluated DU performance when co-hosting AI workloads on GPUs. Since GPUs are already employed to accelerate PHY-layer processing, the authors first showed that conventional resource allocation algorithms fail to effectively coordinate GPU sharing between DU and machine learning workloads on the same host. Building on a detailed analysis of DU GPU usage, they proposed a new resource allocation algorithm that enables ML workloads and DU PHY processing to coexist on the GPU while preserving the DU’s performance requirements. However, these works focus on RAN components themselves, not on real-time dApps.


\noindent{\textbf{dApp Implementations.}} Several works have demonstrated dApp feasibility. D'Oro et al. \cite{d2022dapps} proposed the dApp paradigm. Lacava et al. \cite{lacava2025dapps} implemented spectrum sensing dApps achieving sub-1\,ms latency on bare-metal, with no evaluation on a containerized environment. Olimpieri et al. \cite{olimpieri2025libiq} introduced LibIQ, a library designed to support dApp-based spectrum sensing algorithms. Their evaluation included multiple spectrum sensing methods—both traditional and machine learning-based—and showed that all tested algorithms could execute within real-time limits. Furthermore, Santhi et al. \cite{santhi2025interfo} proposed InterfO-RAN, a dApp-based solution for uplink interference detection in O-RANm leveraging GPU acceleration to achieve inference below the 10\,ms latency requirement. Practical O-RAN experimentation platforms further support the ecosystem \cite{salvat2023open,johnson2022nexran,ko2024edgeric}. All use bare-metal deployments, violating O-RAN's containerization mandate \cite{o-ran2024dapps}.


\noindent{\textbf{Smart NIC Acceleration.}} Smart NICs have been explored for networking tasks such as load balancing \cite{tajbakhsh2022accelerator,tajbakhsh2024p4hauler} and crypto acceleration \cite{zhao2023dis}. Xing et al. \cite{xing2022towards} characterized smart NIC performance, finding that datapath positioning matters more than CPU speed. Panda et al. \cite{panda2022synergy} used smart NICs for 5G dataplane acceleration. Kfoury et al. \cite{kfoury2024comprehensive} provided a comprehensive smart NIC survey; related efforts explore FPGA-based smart NICs and accelerator-aware load balancing \cite{borromeo2022fpga,tajbakhsh2022accelerator}. However, no prior work has evaluated smart NICs for dApp offloading.

\noindent{\textbf{Containerization Overhead.}} Container performance has been studied for general workloads, motivating alternatives such as unikernels \cite{madhavapeddy2013unikernels} and microkernels \cite{liedtke1996toward}. However, dApps differ from typical containerized microservices in three respects: (1)~they operate under a hard 10\,ms deadline where 0.5\,ms of added overhead already consumes 5\% of the budget; (2)~they share CPU cores and the E3 protocol stack with the DU, creating contention absent in isolated microservice deployments; and (3)~they must be co-located with the base station, making deployment placement a first-order design constraint. Consequently, prior general-purpose container benchmarks do not transfer to dApp deployment planning.

\noindent{\textbf{Gap.}} While virtualization and modularity using containers are mandated by O-RAN specifications for fault isolation and resource sandboxing, their impact on real-time dApp latency remains unexplored. The tradeoff between container isolation (separated containers with virtual network interfaces) and performance (co-located containers sharing the host network stack) has not been quantified for sub-10\,ms workloads. Furthermore, multi-dApp scalability under containerization, where statistical multiplexing fails due to hard real-time deadlines, has not been characterized.
No existing work evaluates: (1) O-RAN-mandated containerized dApp deployments, (2) multi-dApp scalability limits, or (3) smart NIC offloading for dApps. Our work addresses all three gaps with a systematic evaluation across four representative workloads and four deployment scenarios.

\section{Setup and Evaluation} \label{sec:methodology}
This section describes the configurations utilized in our testing as well as details our implementation changes to adapt the dApp architecture to work in a smart NIC.

\textbf{Testbed.} Our testbed comprises two servers: (1) RAN Server: Intel Xeon Silver 4210R (10 cores @ 2.4 GHz), 64 GB RAM, NVIDIA RTX 8000 GPU, BlueField-3 smart NIC (16 ARM cores @ 2.6 GHz, 16 GB memory). (2) UE Server: Intel i7-9700 (8 cores @ 3.0 GHz), 32 GB RAM. We use OpenAirInterface v2.3.0 \cite{oairepo} with E3 interface support, operating in TDD mode (20 MHz @ 3.5 GHz), 30\,kHz sub-carrier spacing (numerology~1), and a proportional-fair MAC scheduler. Both servers are connected via Ethernet with the smart NIC inline between network and RAN Server; there is no over-the-air wireless segment, so HARQ retransmissions and RF-induced jitter are excluded by design. We use the OpenAirInterface radio frequency simulator for the UE--gNB link. This provides a controlled baseline that isolates dApp deployment overhead from RF variability. In production, real RF processing (channel estimation, equalization, HARQ) could increase DU CPU load, which would amplify the CPU bottleneck reported in Section~\ref{sec:evaluation}, but we expect the general trends in performance to remain similar.


\noindent{\textbf{Workloads.}} We select four representative dApps: (1) EBS: threshold-based energy detection for spectrum occupancy. (2) FFT: standard DFT for frequency-domain analysis. (3) FCN: 3-layer neural network (512-256-128 neurons, ReLU) trained on DeepSense dataset~\cite{uvaydov2021deepsense} for intelligent spectrum sensing. (4) Xception: 70+ layer depthwise separable CNN \cite{chollet2017xception} for RF fingerprinting, optimized with TensorRT~\cite{NVIDIA2024tensorrt} (GPU) or LiteRT~\cite{google2025litert} (CPU). Default input: 1536 I/Q samples. 
Table~\ref{tab:dapp-selection} summarizes the selection rationale.

\noindent{\textbf{Deployment Scenarios.}} (1) \textit{Bare-metal}: dApp+RAN on host (non-compliant baseline). (2) \textit{Co-located container}: single Docker container for dApp+RAN. (3) \textit{Separated containers}: isolated via virtual network interface. 
(4) smart NIC: entire dApp runs on the BlueField-3's 16 ARM cores (2.6 GHz, 16 GB memory), which sits inline between the RU and RAN node. The dApp directly intercepts  I/Q samples from eCPRI frames as they arrive, eliminating the need to communicate  with the RAN node or main CPU, thereby bypassing E3 protocol overhead. The E3  interface adaptation for smart NIC deployment is detailed below.

\begin{table}[!htbp]
    \centering
    \footnotesize
    \caption{The chosen dApps represent a distinct computational class expected in real-time O-RAN.}
    \begin{tabular}{|l|l|l|l|}
        \hline
        \textbf{dApp} & \textbf{Pattern} & \textbf{Use Case} & \textbf{Represents} \\
        \hline
        EBS & Statistical & Spectrum occupancy & Low-complexity \\
        FFT & Transform & Frequency analysis & Signal processing \\
        FCN & Dense NN & Intelligent sensing & Lightweight ML \\
        Xception & Deep CNN & RF fingerprinting & Heavy ML + GPU \\
        \hline
    \end{tabular}
    \label{tab:dapp-selection}
\end{table}

\begin{figure}[]
\centering
\begin{subfigure}{0.5\textwidth}
  \centering
  \includegraphics[width=0.95\linewidth]{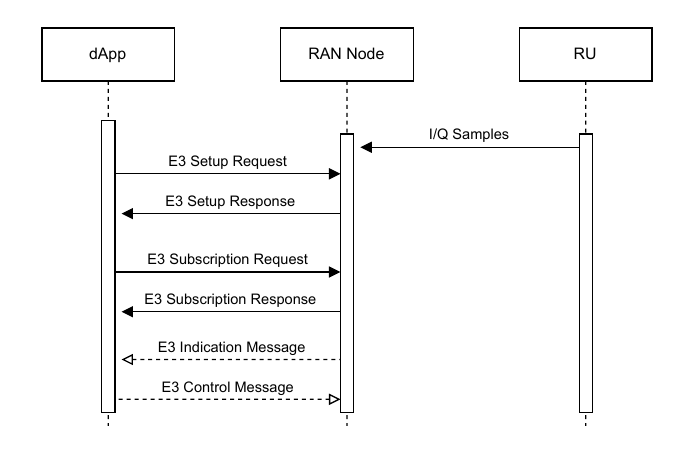}
  \caption{Standard E3 Interface}
  \label{fig:sub1}
\end{subfigure}
\begin{subfigure}{0.5\textwidth}
  \centering
  \includegraphics[width=0.95\linewidth]{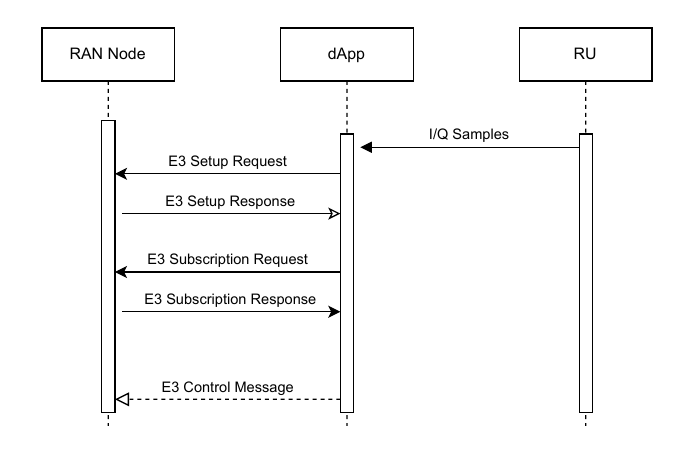}
  \caption{Adapted E3 Interface}
  \label{fig:sub2}
\end{subfigure}
\caption{Messages exchange for the control loop while utilizing the (a) standard E3 Interface and (b) the version adapted to work with a smart NIC.}
\label{fig:interface}
\end{figure}


\noindent{\textbf{Smart NIC-Adapted E3 Interface.}} The key innovation in our smart NIC deployment is the adaptation of the E3 interface to exploit datapath proximity. In the standard E3 interface \cite{lacava2025dapps}, dApps communicate with the RAN node through two components: the E3 Application Protocol (E3AP) for setup/authentication and the E3 Subscription Manager (E3SM) for data exchange via Indication Messages (RAN to dApp) and Control Messages (dApp to RAN). Our smart NIC adaptation preserves E3AP for compatibility but replaces E3SM Indication Messages with direct packet capture. Since the smart NIC sits inline between the RU and RAN node, dApps can intercept I/Q samples directly from incoming eCPRI frames without waiting for the RAN node to encode and transmit E3 messages. This eliminates the Collection phase overhead (E3 decode) and reduces PCIe bus traversals. Control Messages remain unchanged, maintaining standards compliance for the control path. To further explain the difference, we outline the traditional and adapted E3 Interface:

\textit{Traditional E3 Interface:} In~\cite{lacava2025dapps}, the authors propose an E3 interface composed of two main components: the E3 Application Protocol (E3AP) and the E3 Subscription Manager that manages the E3 Service Models (E3SMs). When a dApp starts, it triggers a setup procedure using the E3AP, where it authenticates itself and pairs with the RAN node, performing a subscription request. After the communication is set up, the dApp and RAN node communicate through E3SM messages, namely Indication Messages and Control Messages. An Indication Message is sent from the RAN to the dApp and consists of the necessary information for the dApp to process. The Control Message is the response from the dApp, where after processing the data it outputs a policy or control change and communicates it to the RAN Node.

\textit{Smart NIC-Adapted E3 Interface:} The main difference between deploying dApps on the same host as the RAN node and in the smart NIC is the proximity to the datapath. When placing the applications on the NIC, the data incoming to the node flows directly through it, allowing us to capture this information as it arrives instead of depending on the RAN node to send the information directly. In order to reflect this, we have extended the E3 Interface for this use case. The E3AP stays consistent with the original implementation. The dApp still authenticates itself and establishes a communication channel with the RAN Node. The adaptation comes in the E3SM Indication Messages, which are completely replaced by a direct capture and parsing of the packets that arrive at the RAN node. The Control Message also remains the same.


Figure~\ref{fig:interface} depicts the communication patterns of the standard and the adapted E3 interface. It is important to note in the image the difference in the position of the dApp. In the adapted E3 interface, the dApp receives the I/Q samples directly from the RU, placing itself basically in the middle between the RU and the RAN Node.

\noindent{\textbf{Metrics.}} We measure latency in 4 phases: 1. Collection (decode messages arriving from RAN Node); 2. Processing (inference/computation); 3. Create Control (policy decision); 4. Deliver Control (encode message to be sent to RAN Node). Total latency = sum of phases + network RTT. Resource metrics: CPU/GPU utilization (sampled at 1 s intervals).
Each experiment runs for 5+ minutes, producing around $\sim$300 processed I/Q samples. We report mean, 95th percentile, and CDFs. Moreover, we represent the 10\,ms deadline as a red dotted line in all plots.

\begin{figure}[!htbp]
    \centering
    \includegraphics[width=0.95\columnwidth]{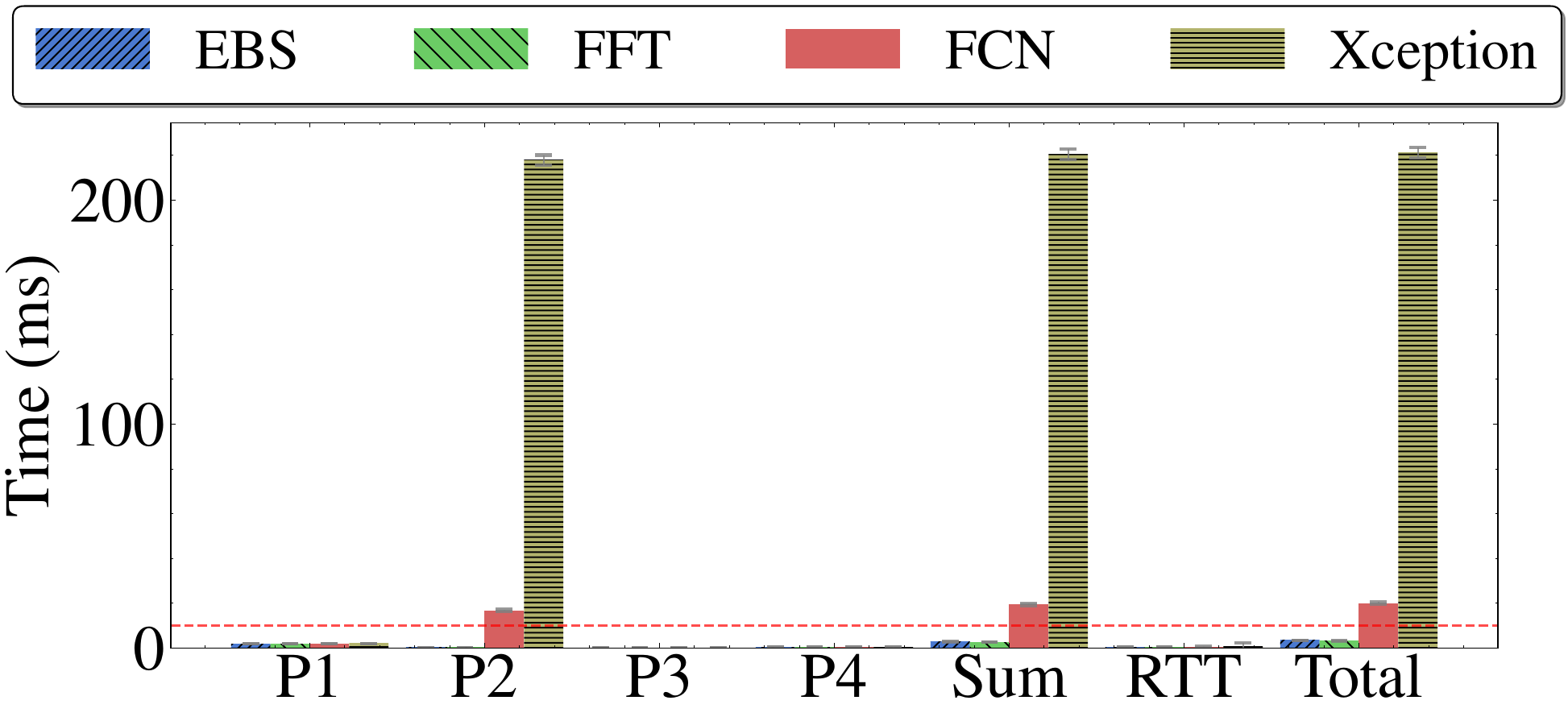}
    \caption{Execution time of dApps on a bare-metal server without optimization. Both FCN and Xception are not able to execute under the threshold without optimizations.}
    \label{fig:feasibility-appendix}
\end{figure}

\begin{figure*}[!htbp]
    \centering
    \includegraphics[width=0.85\textwidth]{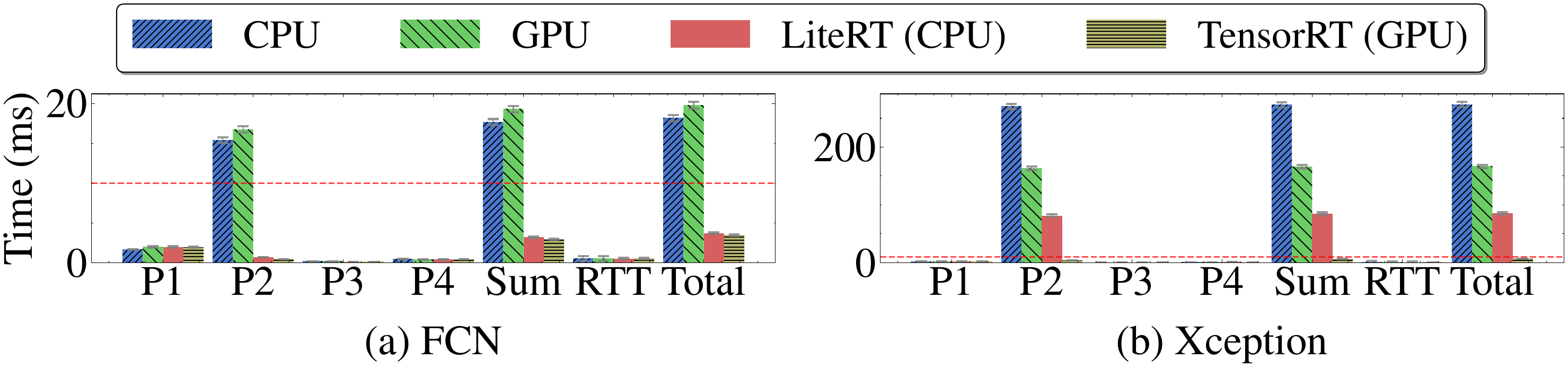}
    \caption{The impact of computing sources and AI platforms on neural network-based dApps in a bare-metal deployment.}
    \label{fig:ml-runtime-appendix}
\end{figure*}

\section{Evaluation Results}
\label{sec:evaluation}

This Section outlines the results of the conducted experiments as well as the main take-aways for each of the proposed research questions.

%


\subsection{Prerequisite: ML Platform Optimization Baseline}

This subsection establishes the optimized configurations used in all subsequent experiments. Unoptimized ML implementations cannot meet the 10\,ms real-time deadline, making optimization strategy selection critical. We replicate baseline measurements from prior work \cite{scalingi2024det} to ensure controlled comparison across our deployment scenarios.

Figure~\ref{fig:feasibility-appendix} shows unoptimized execution times: EBS and FFT meet the 10\,ms bound, whereas FCN ($\sim$18\,ms) and Xception ($\sim$200\,ms) do not. Figure~\ref{fig:ml-runtime-appendix} compares optimization strategies using TensorRT~\cite{NVIDIA2024tensorrt} (GPU) and LiteRT~\cite{google2025litert} (CPU). FCN suffers slight degradation with GPU offloading due to PCIe transfer overhead, while Xception requires TensorRT with GPU to reach sub-10\,ms latency.


\textbf{Optimization Strategy Selection.} For lightweight models (FCN), CPU-optimized frameworks such as LiteRT reduce latency below 10\,ms; GPU offloading degrades FCN performance due to PCIe transfer overhead. Complex models (Xception) require both TensorRT optimization and GPU support to achieve sub-10\,ms latency.

\subsection{RQ1: Can Containers Meet Real-Time Deadlines?}

\begin{figure*}[ht]
    \centering
    \includegraphics[width=0.9\textwidth]{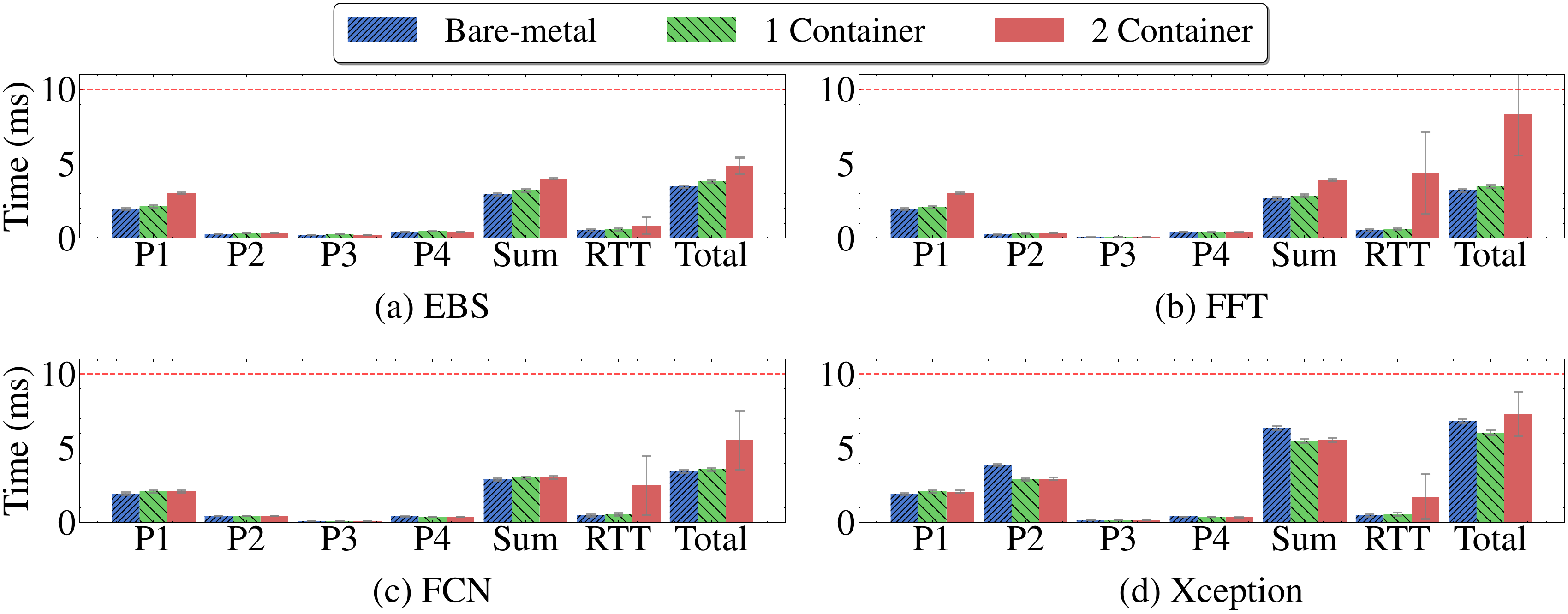}
    \caption{Deployment comparison across four workloads and divided into the 4 latency phases. Co-located containers (2-container deployments) match bare-metal ($<$0.5\,ms overhead), separated containers add 1-2\,ms from virtual networking. Error bars show 95th percentile.}
    \label{fig:dep-env}
\end{figure*}

Figure~\ref{fig:dep-env} compares three deployment options in our workload spectrum, dividing the total time into four phases P1, P2, P3, P4 that correspond  to the latency phases described previously (Collection, Processing, Create Control, Deliver Control). \textit{Co-located containers} (dApp+RAN in one container) achieve near-identical performance to bare-metal with less than 0.5\,ms overhead for all workloads. This shows that the containerization overhead itself is negligible. The Docker runtime and cgroups resource isolation introduce minimal latency when components share the same network namespace. 

On the other hand, \textit{separated containers} (isolated via the virtual network interface) show a different behavior. The virtual network bridge introduces 1-2\,ms of additional latency from packet copying and context switching between network namespaces. For lightweight workloads (EBS, FFT), this overhead is tolerable, as the mean latency remains below 4\,ms. However, for complex workloads, the impact is critical: Xception's mean latency reaches 7.31\,ms in separated containers versus 6.05 ms in co-located containers, close to the 10\,ms deadline.
This reveals a fundamental tradeoff: co-located containers sacrifice fault isolation (a crash in the dApp can affect the RAN node) for performance, while separated containers provide stronger isolation at the cost of potential deadline violations. Operators must choose a deployment model based on their reliability requirements.
For FCN, bare-metal P4 (Deliver Control) is marginally higher than in container deployments ($<$0.3\,ms), attributable to less deterministic CPU scheduling on the uncontainerized host; cgroups enforce stricter time-slicing.

\begin{figure*}[!t]
    \centering
    \includegraphics[width=0.8\textwidth]{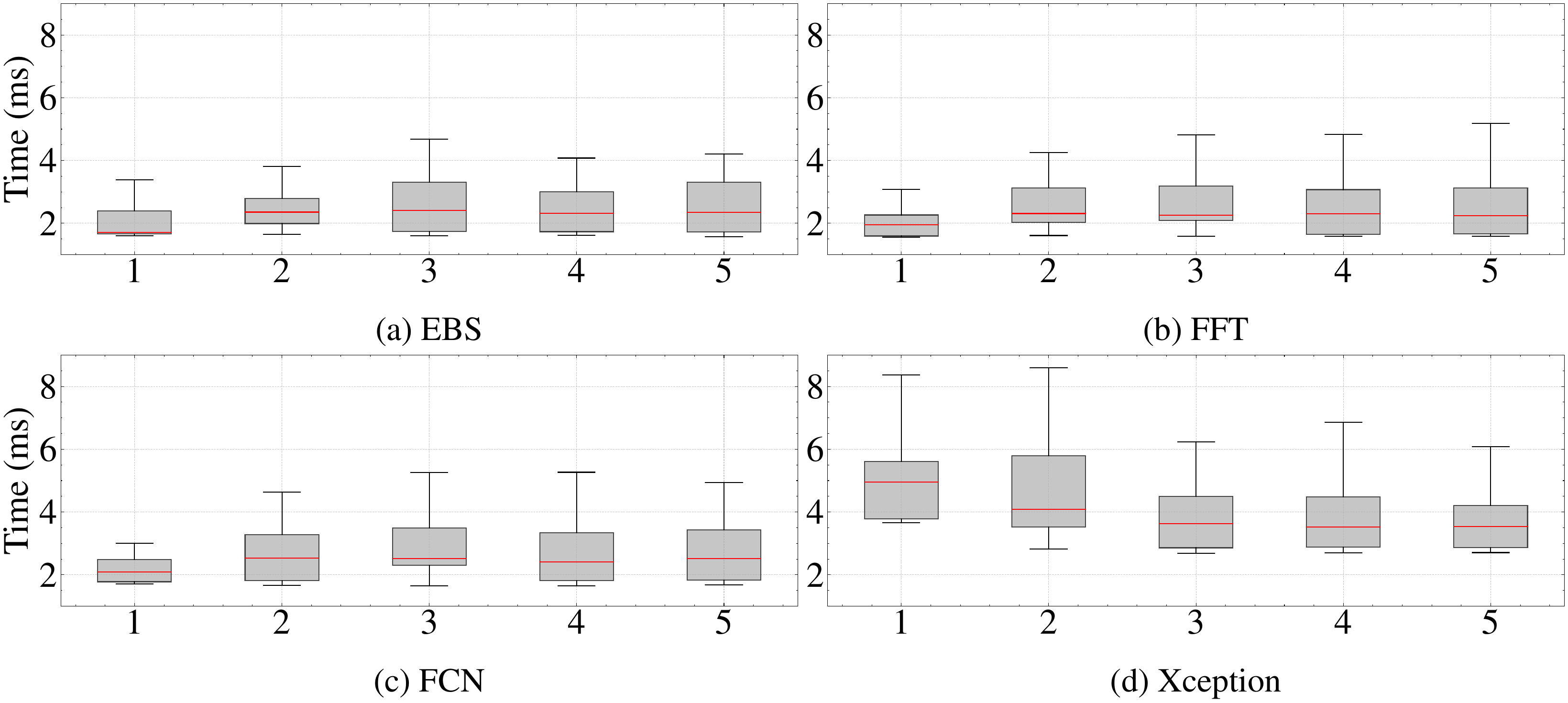}
    \caption{dApp latency vs. allocated CPU cores in a 2-container deployment. Only Xception benefits from multi-core allocation; lightweight dApps degrade due to process migration.}
    \label{fig:cpuset}
\end{figure*}

\noindent{\textbf{Resource allocation.}} 
In the previous experiments, we allocated all ten available CPU cores to both the dApps and the RAN node. However, the CPU demand of a dApp often depends on its computational complexity. Thus, in this evaluation, we examine the per-dApp CPU requirements by deploying each dApp and the RAN node in separate containers. Figure~\ref{fig:cpuset} shows the latency of the four dApps as the number of allocated CPU cores varies. The results indicate that only the Xception-based dApp benefits from additional CPU cores. For the remaining dApps, allocating more than one core unexpectedly degrades performance. 
This behavior stems from the fact that 3 dApps (except for Xception) are single threaded applications and do not gain any benefits from multiple cores. Instead, the OS scheduler migrates these processes across cores, forcing expensive memory accesses and context migrations.
In contrast, Xception dApp is a multi-threaded application and adding additional cores expedites its preprocessing before conducting the inference using GPU. 

\noindent{\textbf{Takeaway.}} 
Operators should consider co-located containers (i.e., deploying the dApp and RAN components within a single container) for latency-critical dApps where sub-10\,ms deadlines are mandatory. The negligible overhead ($<$0.5\,ms) makes this configuration the preferred choice for performance-sensitive deployments. On the other hand, separated containers should be used only when strong fault isolation is required, as this design introduces an additional 1–2\,ms of virtual networking latency. For example, using separated containers raises the mean latency of Xception to 7.31\,ms, leaving very little headroom before violating real-time constraints.

Computing resource allocation must also account for (a) the computational complexity and (b) the multi-threading characteristics of deployed dApps. Lightweight, memory-bound dApps (e.g., EBS, FFT, FCN) should be pinned to a single core using \texttt{cpuset} to preserve L1 cache locality. Assigning additional cores to these workloads degrades performance by 15–30\% due to cache invalidation caused by OS-level core migration. In contrast, only compute-intensive models with multi-threaded implementations, such as Xception, benefit from multi-core allocation.

\subsection{RQ2: What Limits Multi-dApp Scalability?}

It is common to have multiple instances of a dApp running concurrently to serve different operators. Thus, the focus of this section is on characterizing the computational resource usage of dApps running concurrently. Specifically, we deploy 2-16 concurrent dApps in separated containers (the strictest O-RAN-compliant configuration) on our 10-core CPU to understand scalability limits. This range allows us to observe both pre-saturation (2-10 dApps) and post-saturation (11-16 dApps) behaviour. All dApps run the same workload, to allow us to isolate the impact of scale.

\begin{figure}[ht]
\centering
  \centering
  \includegraphics[width=.9\columnwidth]{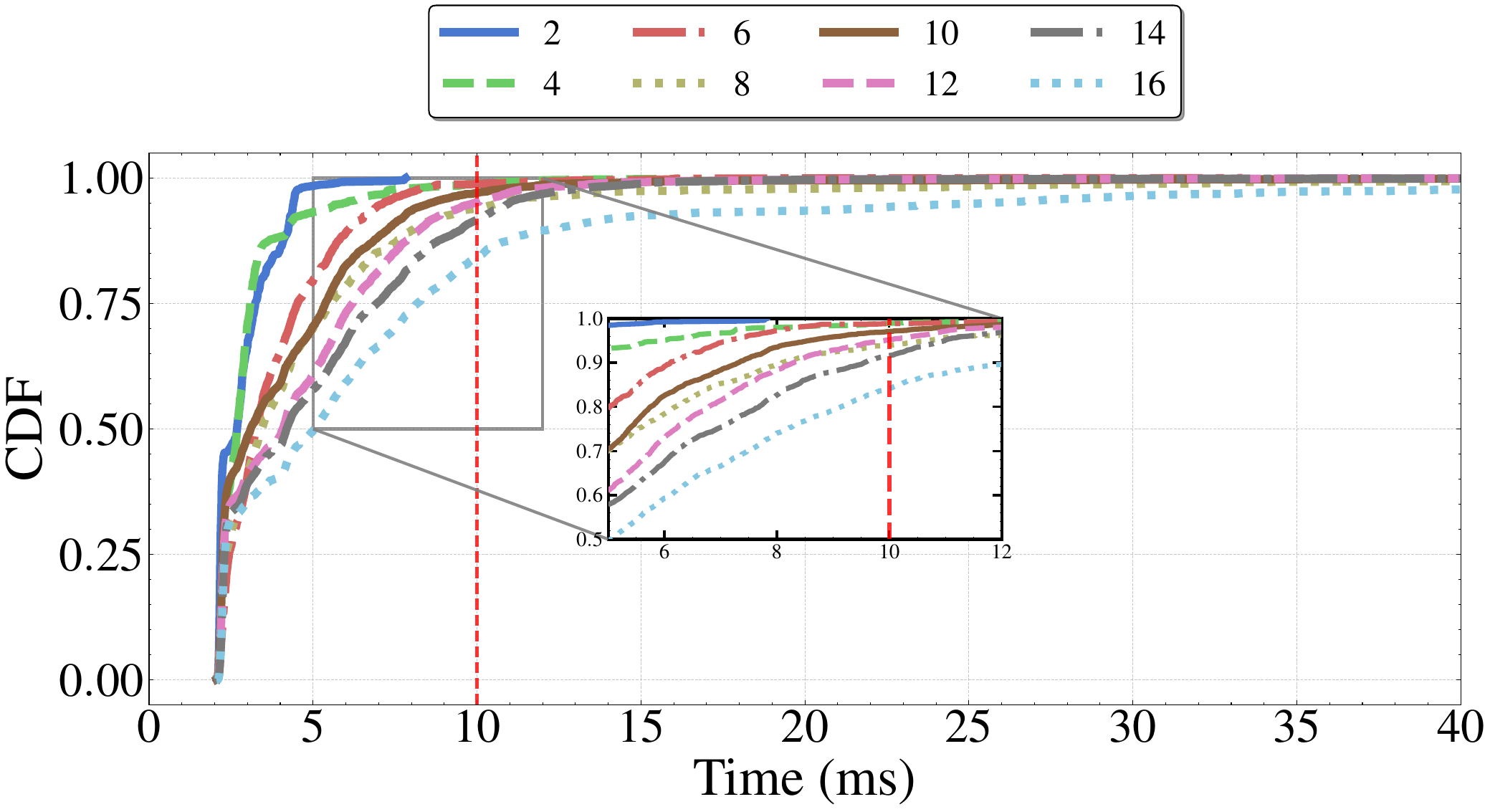}
\caption{CPU-based FCN running at up to 16 instances simultaneously in a container deployment. Sharp increase in latency when deploying more dApps than cores available (10).}
\label{fig:contention-cpu}
\end{figure}

Figure \ref{fig:contention-cpu} presents the CDF of the FCN latency. 
Performance scales well up to 6 dApps, with latencies consistently below 7\,ms. Beyond this point, degradation becomes visible: with 8 dApps, the median latency increases to approximately 8\,ms and with 10 dApps (matching the available core count), the latency reaches 9-10\,ms with approximately 10\% samples exceeding the deadline. At 16 dApps, performance severely degrades, with around 20\% of samples violating the deadline. In contrast to cloud services where CPUs can be oversubscribed by 2x-4x, real time dApps cannot tolerate scheduling delays. When dApps exceed the available cores, the OS scheduler introduces unpredictable latency spikes as processes queue for production.


\begin{figure}[ht]
\centering
  \centering
  \includegraphics[width=.95\columnwidth]{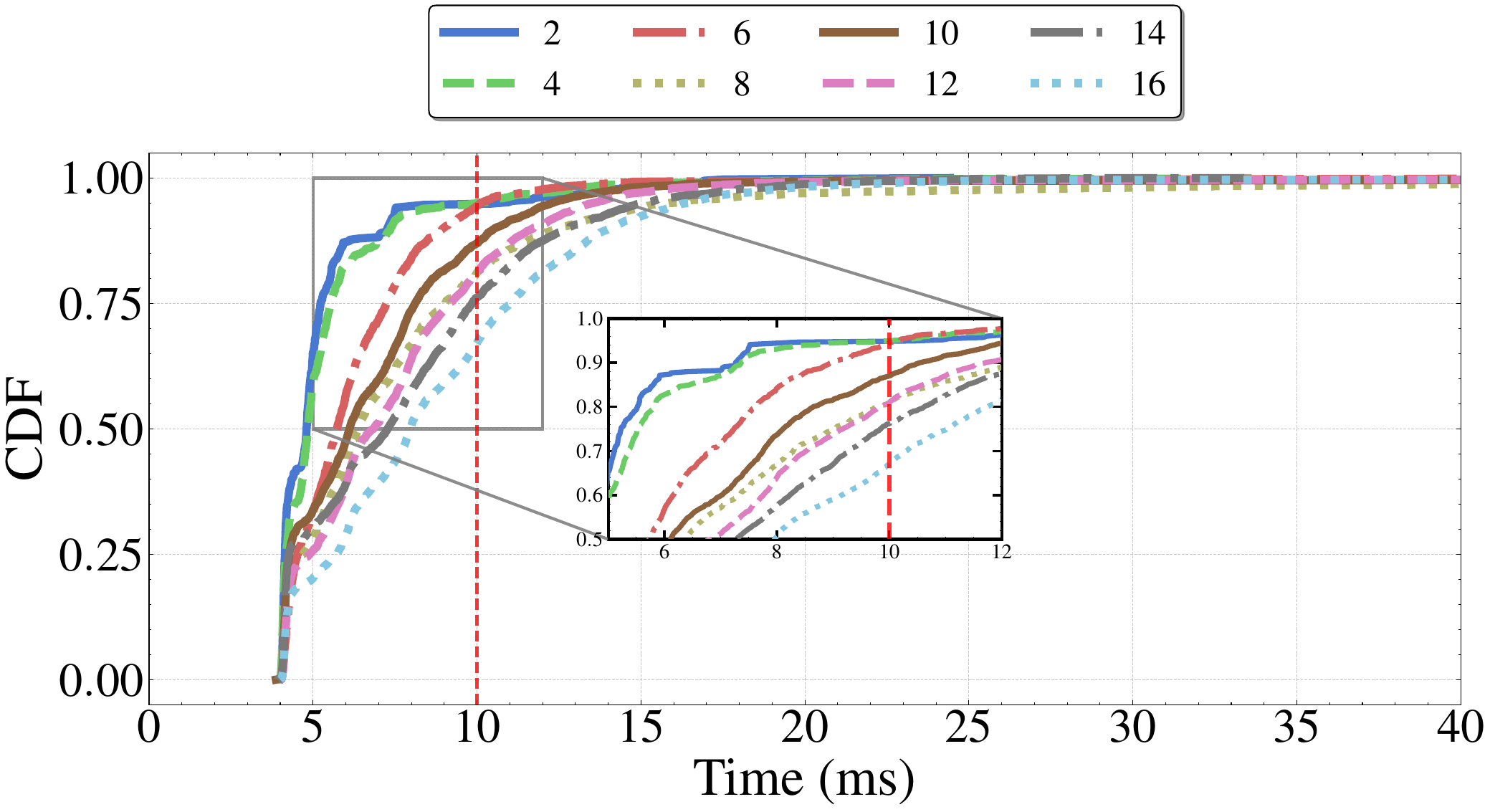}
\caption{GPU-accelerated Xception at up to 16 instances. Around 40\% of samples violate 10\,ms deadline when deploying the max number of applications.}
\label{fig:contention}
\end{figure}

\begin{figure}[ht]
    \centering
    \includegraphics[width=0.95\columnwidth]{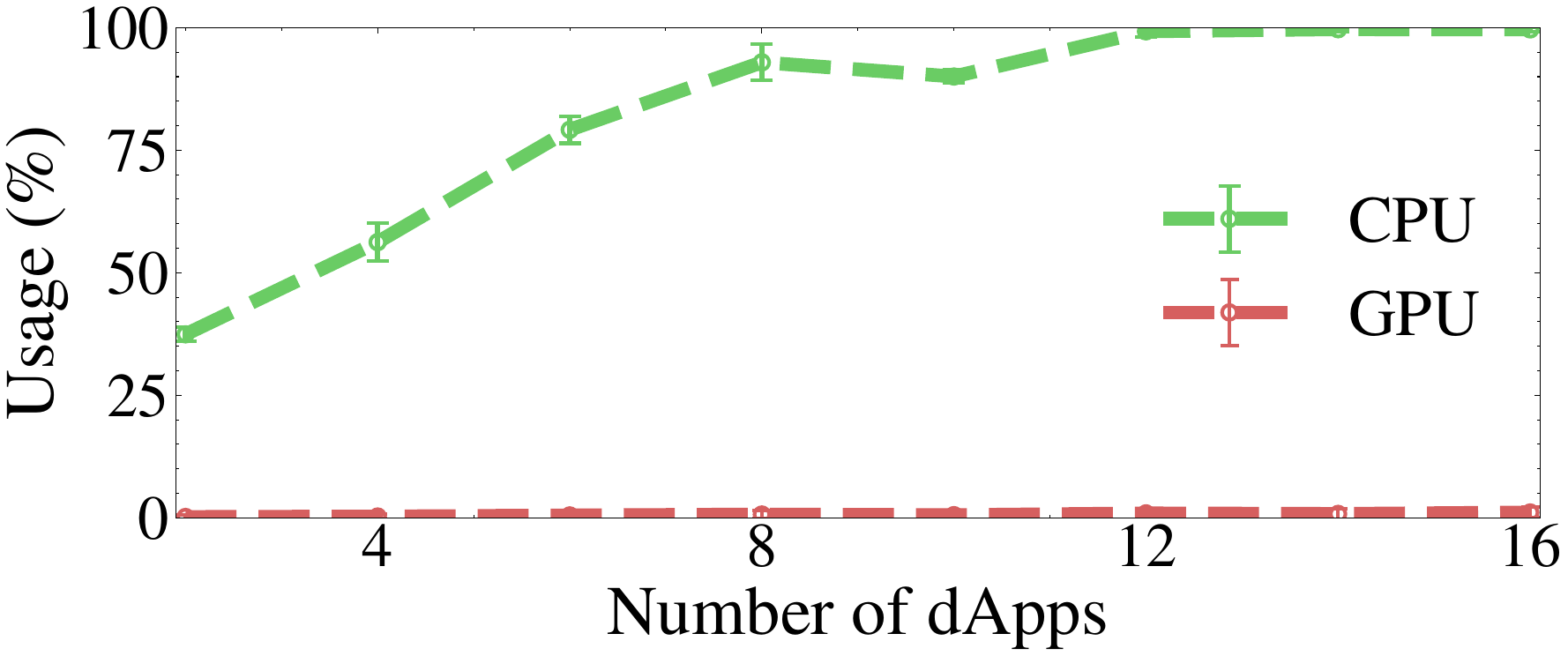}
    \caption{Resource utilization reveals the bottleneck: GPU idle ($<$10\%) while CPU saturates (100\%).}
    \label{fig:metrics}
\end{figure}

The next evaluation examines the scalability of concurrent Xception dApp deployment. 
Figure~\ref{fig:cpuset} already demonstrated that Xception benefits from multi-core CPU allocation due to its preprocessing demands, suggesting that CPU availability would be critical for scalability. Indeed, Figure~\ref{fig:contention} confirms this concern. When running 16 concurrent Xception dApps (optimized with TensorRT) across all CPU cores and the GPU, the CDF shows that while 60\% of samples meet the 10\,ms deadline, 40\% violate it, with the 95th percentile reaching approximately 12\,ms and the 99th percentile extending to 15\,ms. This motivated a deeper analysis, presented in Figure~\ref{fig:metrics}.

The results show that GPU utilization remains below 2\% even with 16 dApps, while the CPU becomes fully saturated. This imbalance arises from the Xception pre-processing pipeline (parsing eCPRI frames, extracting I/Q samples, and performing input normalization and reshaping), which consumes 3-4\,ms per dApp on the CPU despite inference being offloaded to the GPU. The GPU completes inference in under 2\,ms but spends most of its time idle, waiting for the CPU to prepare the next input.

\noindent{\textbf{Takeaway.}} 
These results demonstrate that edge platforms cannot rely on statistical multiplexing for real-time workloads; capacity planning must be based on discrete CPU allocation. GPU acceleration alone is insufficient because system-level bottlenecks (particularly CPU-bound preprocessing and protocol handling) may dominate end-to-end latency. Hence, CPU availability may become the limiting factor in multi-tenant deployments. Thus, operators should provision roughly one CPU core per dApp or distribute workloads across multiple hosts before reaching saturation. Moreover, future O-RAN hardware designs should prioritize higher CPU core counts over additional GPU compute for scalable, real-time dApp execution, especially if the workload is CPU bound. 



\subsection{RQ3: Can Smart NICs Address Bottlenecks?}

\begin{figure}[!htbp]
    \centering
    \includegraphics[width=0.95\columnwidth]{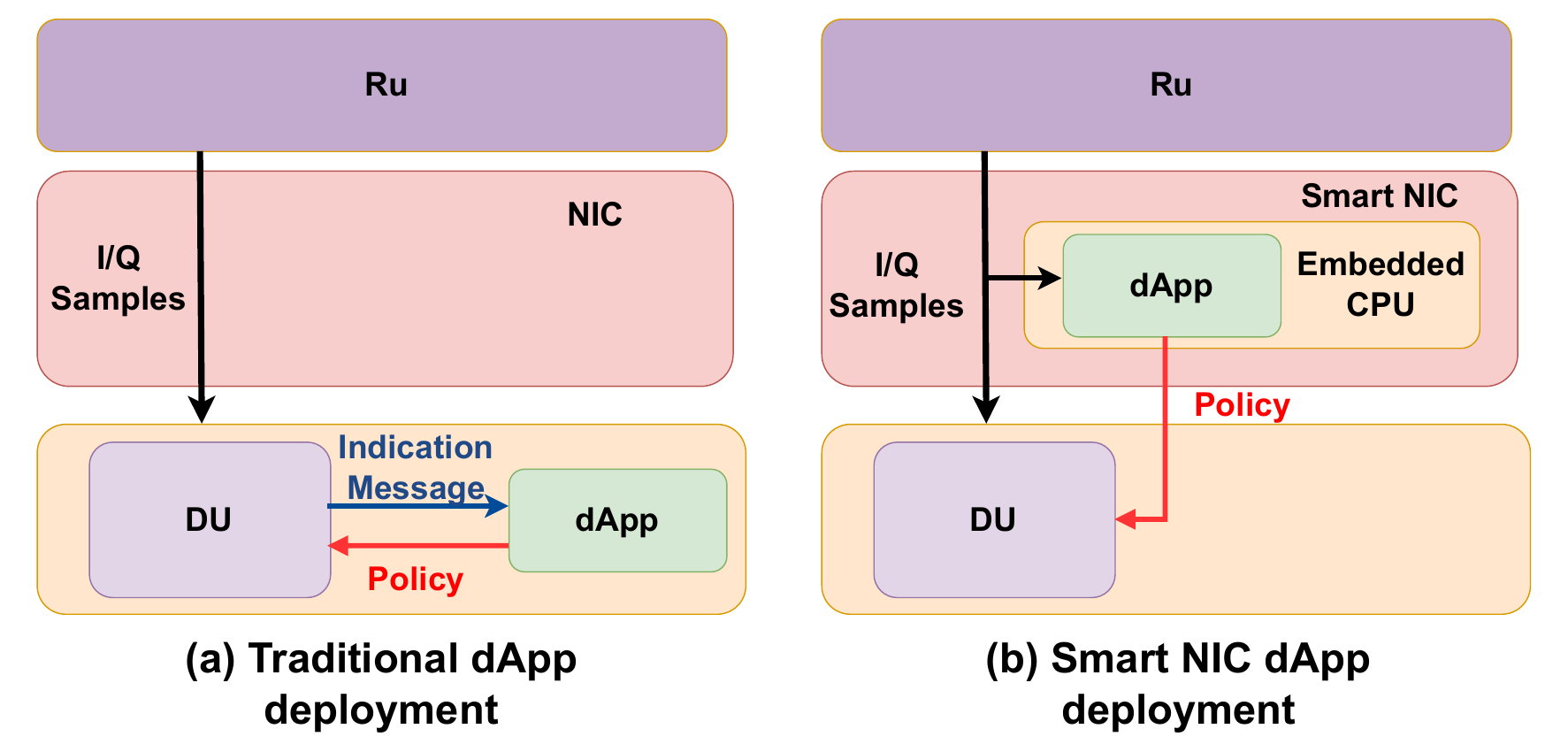}
    \caption{Comparison between traditional dApp deployment and smart NIC based one.}
    \label{fig:smartnic-arch}
\end{figure}

RQ2 revealed two key insights: (1) simple dApps can be efficiently hosted on standard CPU cores, and (2) CPU cores can become bottlenecks that degrade the performance of concurrent dApps. Thus, this research question focuses on evaluating the potential of emerging programmable smart NICs to reduce CPU contention by judiciously selecting and offloading suitable dApps to run on these devices.
Smart NIC offloading complements rather than replaces containerization: lightweight dApps execute on the smart NIC while GPU-dependent dApps remain containerized on the host.
As the smart NIC sits on the datapath between the fronthaul and the RAN node, as can be seen in Figure~\ref{fig:smartnic-arch}, dApps running on the smart NIC can directly intercept and process incoming I/Q samples without involving the RAN node or the main CPU. Thus, we begin by characterizing the reference performance of dApps offloaded to the smart NIC and compare these results against a bare-metal server to evaluate deployment feasibility and achievable latency. 

\begin{figure*}[!htbp]
    \centering
    \includegraphics[width=0.99\textwidth]{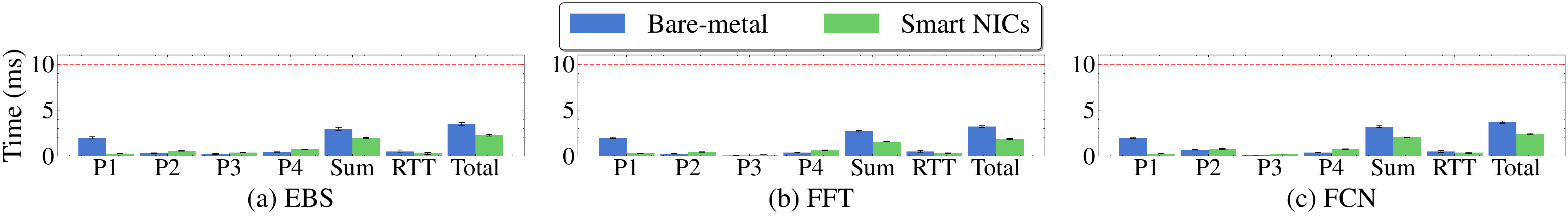}
    \caption{Smart NIC deployment outperforms bare-metal in terms of latency despite slower CPUs (2.6 GHz ARM vs. 3.2 GHz Xeon) with all three dApps. Datapath positioning eliminates E3 protocol overhead.}
    \label{fig:snic}
\end{figure*}

Figure~\ref{fig:smartnic-arch} illustrates the architectural difference: in smart NIC deployment, the dApp intercepts I/Q samples directly from the datapath, bypassing E3 protocol overhead. All three lightweight dApps (EBS, FFT, and FCN) meet the 10\,ms requirement on the smart NIC. 
Figure~\ref{fig:snic} confirms that smart NIC deployment achieves lower latency than bare-metal for all three dApps, as it captures I/Q samples directly from the traffic flow, bypassing E3 encoding and PCIe bus traversal.



\begin{figure}[h]
        \centering
    \includegraphics[width=0.95\columnwidth]{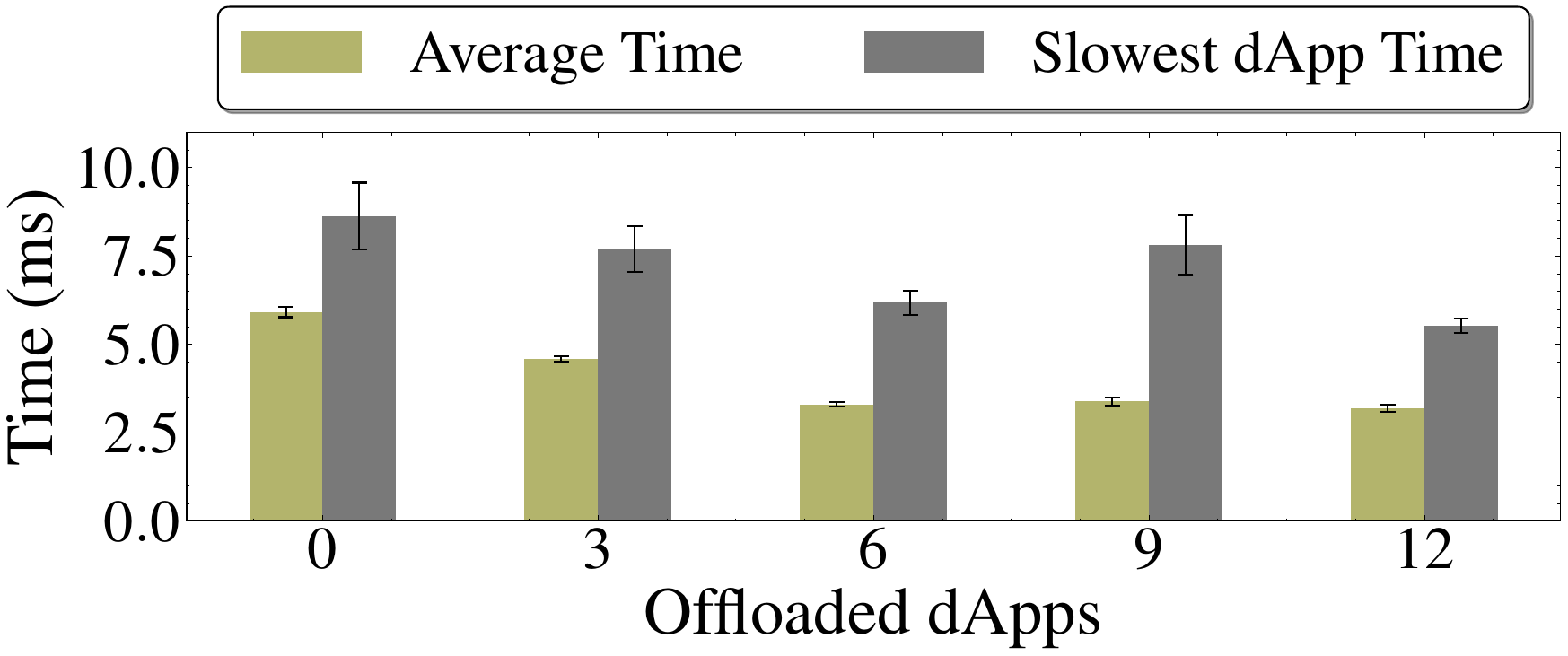}
    \caption{The average response time reduces by 35\%, worst-case latency by 42\%, and deadline violations from 18\% to 3\% when offloading lightweight dApps (EBS, FFT, FCN) to the smart NIC.}
    \label{fig:offloading}
\end{figure}

In the second evaluation, we examine the scalability of NIC-based dApp deployment by incrementally offloading lightweight dApps from the server to the smart NIC. We start with 16 dApps on the server (4 instances each of EBS, FFT, FCN, and Xception) and progressively offload 3, 6, 9, and 12 lightweight dApps to the smart NIC in the following manner: first we offload one of each EBS, FFT, and FCN, then two of each and so on. Xception remains on the server throughout due to its GPU requirements. Figure~\ref{fig:offloading} shows that as we offload more dApps, both average processing time and worst-case latency (representing the slowest dApp) improve. 

\noindent{\textbf{Takeaway.}} 
Smart NICs achieve lower latency than bare-metal servers, despite using slower ARM cores (2.6 GHz vs. 3.2 GHz Xeon). This highlights that architectural placement (embedding compute directly in the datapath to eliminate protocol overhead) can outweigh raw CPU speed for memory-bound workloads. Smart NICs remove packet encoding/decoding overhead, reduce host CPU load, and address critical bottlenecks by positioning additional processing cores inline with packet flow, which traditional servers cannot avoid.

\subsection{Discussion}

Our experiments define a complete set of evaluations that establish the performance of dApps over multiple deployments and a variety of parameters. O-RAN operators can leverage the findings of this work to guide further deployment of dApps. However, there are some key aspects that could be explored in possible extensions.

\textbf{Real Radio Link.} Our current evaluation relies on standard virtualized implementations of RAN nodes and a realistic Radio Frequency simulator. However, real RF propagation may alter RAN CPU load due to channel estimation and equalization, potentially worsening the CPU bottleneck. Extending this work to encompass over-the-air experiments would further improve the result by quantifying how realistic RF conditions affect resource provisioning and whether smart NIC offloading becomes more critical in production. Since the RF simulator excludes HARQ retransmissions and channel-induced jitter, the reported latencies constitute a lower bound; production RF conditions would increase DU CPU utilization and amplify the bottleneck identified in RQ2.

\textbf{Deployment Scale.} Our experiments delved into the impact of scale in terms of number of concurrent dApps. However, further evaluation in terms of larger RAN deployments, considering multiple DUs, RUs, and connected smart NICs can further improve results, outlining the impact of network scale on dApp performance. Moreover, expanding the set of dApps tested with options that receive other input types (e.g., buffer status reports, CQI, HARQ feedback) would enable the exploration of mixed-workload deployment strategies.

\section{Conclusion}
\label{sec:conclusion}

We provide the first systematic evaluation of O-RAN-compliant containerized dApp deployments, addressing a critical gap between O-RAN specifications and existing implementations. Our evaluation across four representative workloads (EBS, FFT, FCN, Xception) and four deployment scenarios (bare-metal, co-located containers, separated containers, Smart NIC offloading) reveals key insights that enable operators to balance O-RAN compliance, real-time guarantees, and multi-tenant scalability. Our findings have broader implications for real-time network functions: datapath positioning matters more than CPU speed, and GPU acceleration is necessary but not sufficient when CPU-bound pipeline phases dominate end-to-end latency.
Specifically, containerization is viable for real-time dApps if isolation is carefully managed. Co-located containers match bare-metal performance with negligible overhead ($<$0.5\,ms), while separated containers add 1-2\,ms of networking overhead that can push complex workloads near deadline violations. Next, CPU provisioning must scale linearly with dApp count at a 1:1 ratio due to real-time constraints that prevent statistical multiplexing. Even GPU-accelerated workloads saturate CPU due to E3 protocol parsing and preprocessing, with GPU utilization staying below 10\% while CPU reaches 100\%. Finally, smart NIC datapath integration can be an interesting option for deployments, providing benefits for memory-bound workloads, achieving 15-25\% lower latency despite slower ARM cores and enabling 50\% more dApp capacity per server through heterogeneous offloading.

As part of our future work, we plan to design and implement an orchestration framework for automated dApps placement across host CPUs and smart NICs. This framework will leverage workload characteristics and key insights derived from our deployment measurements to make informed placement decisions. In addition, we intend to explore a holistic resource optimization strategy that utilizes all available hardware resources, enabling independent optimization of each processing stage. 

\bibliographystyle{IEEEtran}
\bibliography{sample-base}

\end{document}